\documentclass[useAMS]{mn2e}
\usepackage{graphicx}
\usepackage{hyperref}

\usepackage{epsfig}
\usepackage{float}
\usepackage{txfonts}
\title[The activity of RU\,Peg]{The activity of the dwarf nova RU\,Pegasi with rapidly changing outburst types}
\author[V.~\v{S}imon]{Vojt\v{e}ch~\v{S}imon$^{1,2}$\thanks{E-mail:simon@asu.cas.cz}\\
$^{1}$Astronomical Institute of the Czech Academy of Sciences, 25165 Ond\v{r}ejov, Czech Republic \\ 
$^{2}$Czech Technical University in Prague, Faculty of Electrical Engineering, 16627 Prague, Czech Republic}
\begin{document}

\date{Received 2021 xxx x}

\pagerange{\pageref{firstpage}--\pageref{lastpage}} \pubyear{2021}

\maketitle

\label{firstpage}

\begin{abstract}
RU\,Peg  is  a dwarf  nova  (DN)  of  the  U\,Gem type. Our  analysis  of its 
long-term  optical  activity  uses  the  data from  the  AAVSO  database.  It 
concentrates on investigating the properties of  the individual outbursts and 
the  time  evolution  of  the  ensemble  of  these   events.  No  significant 
irradiation   of  the  disc  by  the   white  dwarf   was  detected.  In  the 
interpretation, a variable steepness of the rising branches of the individual 
outbursts shows that the start of outbursts of  RU\,Peg  can occur at various 
distances from  the  disc  centre without remarkable changes of the  outburst 
recurrence time $T_{\rm  C}$. The disc overflow of the inflowing mass stream, 
varying with time, could contribute  to the  changes in the starting position 
of the heating  front, hence the variations of the outburst types.  A typical 
length of  $T_{\rm  C}$  was 90\,days. The segments of the  relatively stable 
length of $T_{\rm  C}$  were accompanied by the primarily little variable and 
small values of the fluence (the energy  radiated in the  optical band in the 
individual   outbursts).  Jumps  of  $T_{\rm  C}$,  accompanied  by  the  big 
scatter of the  fluences, sometimes replaced them. In  the  interpretation, a 
combination of variations of $T_{\rm  C}$ with the unstable properties of the 
outbursts, including an  unstable mass  transfer rate between the components, 
shows the influence of several mechanisms on the state of the disc in various 
time segments. 
\end{abstract}


\section{Introduction    \label{int}   }

     Cataclysmic variables  (CVs) are  close  binary  systems in which matter 
transfers onto the white dwarf (WD) from its companion, a  lobe-filling  star 
(the secondary, the donor) (e.g., Warner 1995). Orbital periods $P_{\rm orb}$ 
of CVs are from several minutes to several  days  (Ritter  \&  Kolb 2003, and 
updates). The time-averaged mass transfer rate $\dot m_{\rm  tr}$ between the 
donor and the accretor plays a significant role in governing these CVs' types 
of long-term activity. 

     If the value of $\dot m_{\rm  tr}$ of a given CV is between some limits, 
the accretion  disc  embedding  the  WD is  exposed  to  the  thermal-viscous 
instability (TVI). It leads  to episodic accretion  of matter  from  the disc 
onto the WD. When the  column density  of matter accumulated in the accretion 
disc reaches a critical value, propagating the heating  front across the disc 
starts the outburst. It leads to a quick transition of the disc from the cool 
state  to  the  hot  state.  An  increase  in  its  viscosity accompanies it. 
Substantial accretion of  matter  from the  disc occurs during this outburst. 
The outburst  finished  when  the disc  became  too  depleted by accretion to 
remain in  the hot  (ionized) state. Therefore, the propagation  of a cooling 
front across the disc finishes this outburst (e.g., Smak 1984; Hameury et al. 
1998). Such CVs are called dwarf novae (DNe) (e.g., Warner 1995). The TVI and 
the variations of $\dot m_{\rm tr}$ between the components play a significant 
role in CVs activity. 

     The model of  Hameury  et al.  (2000)  showed  the  combined  effects of 
irradiating the accretion disc  and the donor and evaporating the inner parts 
of the discs of DNe. These  effects influenced the  predicted DN light curves 
significantly. They confirmed the  suggestion by Warner (1998) that the large 
variety of observed  outburst  behaviour may result from these three effects' 
interplay. Hameury  et al. (2000)  also  obtained  long outbursts, similar to 
superoutbursts, without assuming the presence of a tidal instability. 

     The model of  Schreiber  \&  Hessman (1998) shows that the stream impact 
onto the  accretion disc  and its possible  overflow  can alter  heating  and 
cooling fronts' behaviour in the disc with the TVI. The deposition of mass in 
the disc's inner  parts  can  significantly  change  the eruption light curve 
character. 

     The donor's  magnetic  activity  can also  influence  the accretion disc 
structure (Pearson et al. 1997). It removes  angular  momentum  from the disc 
material, increasing the inward mass flow. This makes the accretion disc more 
centrally condensed, causing  a reduction in  the  outbursts' recurrence time 
$T_{\rm C}$. 

     In summary, Dubus et al. (2018)  showed that DNe are consistently placed 
in the unstable region of the  $P_{\rm  orb}$  vs  $\dot  m_{\rm  tr}$ region 
predicted by the  TVI. Nevertheless, Hameury  (2020)  brought  arguments that 
additional free  parameters, e.g., mass  transfer  variations, irradiation of 
the accretion disc, winds, had to be added to the model. 

     RU\,Peg is a DN of the  U\,Gem type  (Howarth 1975;  Samus et al. 2017). 
Its distance $d = 274.38 \pm 3.4$\,pc was determined from the observations by 
the satellite {\it Gaia} (Gaia Collaboration: Brown et al. 2018; Bailer-Jones 
et al. 2018) 
\footnote{\url{http://vizier.cfa.harvard.edu/viz-bin/VizieR?-source=I/347}}. 
Stover  (1981)  determined its  $P_{\rm  orb}$  of 0.3746\,d  and a secondary 
spectral type K2--5V. Friend et al. (1990)  determined  the inclination angle 
of  the  orbital  plane  $i \approx 33^{\circ}$  and  a WD mass of  $1.38 \pm 
0.06$\,M$_{\odot}$. They argued that the secondary component must have a mass 
less than predicted by Patterson's (1984) standard  main-sequence mass-radius 
relations.

     The {\it  IUE} spectra of RU\,Peg, obtained in deep quiescence, showed a 
very hot WD effective temperature $T_{\rm  eff}$  of 50\,000--53,000\,K (Sion 
\& Urban 2002). The {\it  FUSE}  spectrum in quiescence showed an even higher 
value, 70\,000\,K (Godon et al. 2008). 

     Balman et al. (2011) found an X-ray spectrum harder than most  DNe. This 
indirectly  confirmed  the  large  mass  of  the  WD  in  RU\,Peg. The  X-ray 
luminosity corresponded to a BL luminosity for  a mass  accretion  rate of $2 
\times 10^{-11}$\,M$_{\odot}$\,yr$^{-1}$  (assuming  the  mass  of  the WD of 
1.3\,M$_{\odot}$). Dobrotka et al. (2014) argued that accretion from the disc 
occurs via  the boundary layer in quiescence of RU\,Peg. Therefore, it is not 
an intermediate  polar. A  disc  truncation  radius  is at  most  $0.8 \times 
10^{9}$\,cm (Dobrotka et al. 2014). 

    The secondary (donor) of RU\,Peg is highly magnetically active, similarly 
as in CVs BV\,Cen and AE\,Aqr. The Roche tomograms show  a near-polar spot on 
this component (Dunford et  al. 2012). These tomograms  of RU\,Peg  also show 
prominent irradiation of the secondary star's hemisphere  confined to regions 
that directly  view the accretion regions near the peak magnitude of outburst 
(Dunford et al. 2012). 

    In this paper, we  investigate  the  evolution  of the  long-term optical 
activity of RU\,Peg. A preliminary version of this analysis  was presented by 
\v{S}imon (2018b, 2019).

\section{Observations}      \label{obs}   

                   The  AAVSO  International  database  (Massachusetts,  USA) 
\footnote{\url{https://www.aavso.org/data-download}}  (Kafka  2019)  contains 
both the visual and CCD measurements of RU\,Peg. If treated carefully, visual 
data can be beneficial for  analyzing  the long-term  activity of the objects 
with  a large  amplitude  of  the  changes  of brightness (Percy et al. 1985; 
Richman et al. 1994). Even accuracy  better than 0.1\,mag can  be achieved by 
averaging these data. It  is quite sufficient  for  analyzing large-amplitude 
variable objects like RU\,Peg (several magnitudes).

\begin{figure}
\centering
 \includegraphics[width=0.49\textwidth]{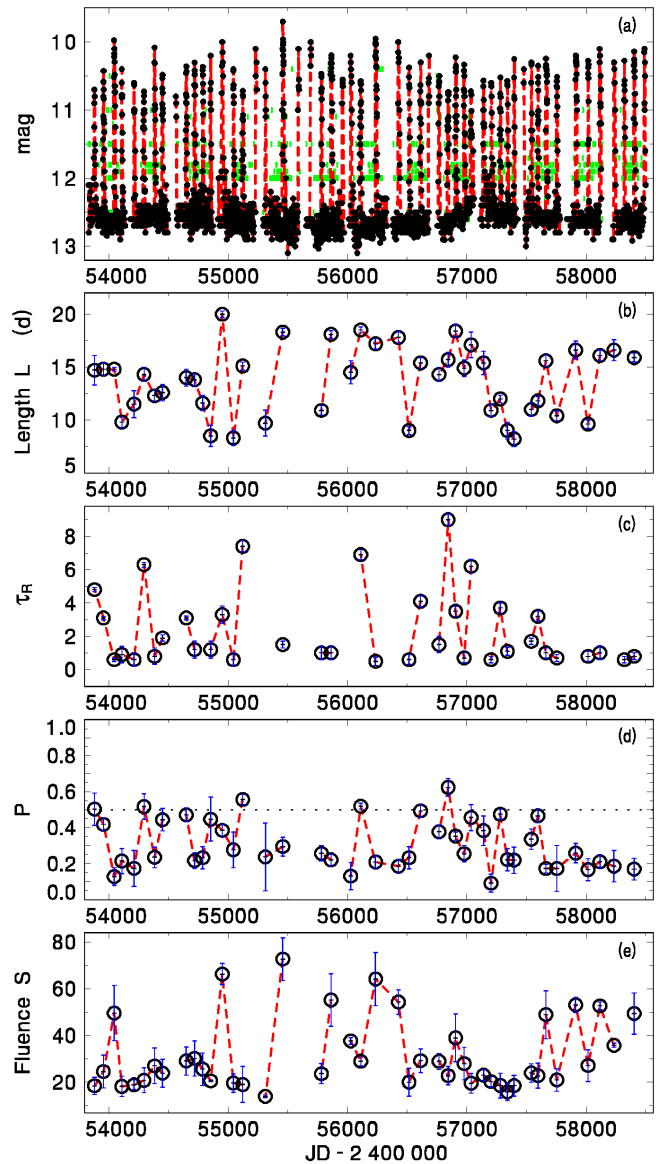} 
\caption{{\bf (a)} The long-term activity of  RU\,Peg. The points 
denote the 1-day means of  brightness. In the  densely  populated 
segments, they are connected by the line to  guide  the  eye. The 
standard deviations of brightness are  marked. The short vertical 
lines denote  the  upper limits  of  brightness.  {\bf  (b)}  The 
length of  the  outburst  measured  at the  brightness 12.2\,mag. 
{\bf  (c)}  The  steepness  of the rising branch of the outburst, 
$\tau_{\rm R}$, in d\,mag$^{-1}$.  {\bf  (d)}  Time  evolution of 
parameter $P$ (Eq.\ref{eq1}) of the outburst. The horizontal line 
denotes  a symmetric  outburst.  {\bf  (e)}  Fluences $S$  of the 
outbursts.  Dimensionless   units  were   used.  The  neighboring 
well-covered outbursts in {\bf  (b)}, {\bf  (c)}, {\bf  (d)}  and 
{\bf  (e)} are connected by a line to guide the eye. The standard 
deviations are marked. They are  comparable  to  the  size of the 
symbol in some cases. See Sect.\,\ref{ana} for details.   } 
\label{ru-jd}
\end{figure}

     The band of  the  sensitivity  of  the  visual  data  is  similar to the 
$V$-band. To be compatible with the optical band, we  used only  the $V$-band 
and  non-filtered  CCD  data.  We  checked  the  reliability  of  the  visual 
measurements by comparing the visual and CCD observations of RU\,Peg included 
in the AAVSO  database. An inspection  of the  light  curve  showed  that the 
magnitudes determined by these two methods were in good mutual agreement.

\section{Data analysis}    \label{ana}   

     The light curve was plotted, and each outburst was  carefully inspected. 
Because this  analysis  focuses on  the  long-term  activity  of RU\,Peg, the 
arithmetic means of these data  were  calculated. Therefore, a 1-day  mean of 
brightness was determined from an ensemble of the visual and CCD measurements. 
We found that  98\,per\,cent of these 1-day means  of brightness had an error 
less than 0.3\,mag. A typical error was 0.15\,mag.

\subsection{Overview of the activity  \label{ove} }  

     A series of DN outbursts of RU\,Peg is displayed in  Fig.\,\ref{ru-jd}a. 
Only the well-covered  part of the light curve was  selected. The coverage by 
the detections is dense, with only the short seasonal gaps. 

\begin{figure}
\centering
 \includegraphics[width=0.47\textwidth]{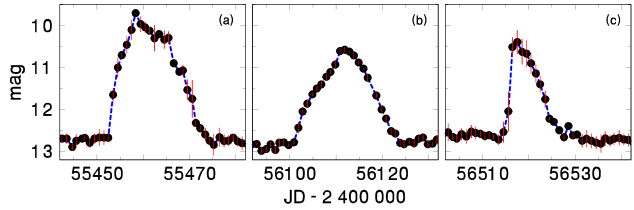} 
\caption{Examples of light curves of outbursts in RU\,Peg. 
{\bf (a)} Bright outburst with a plateau on the top. {\bf 
(b)} Outburst with a very long (slow) rising branch and a 
sharp peak. {\bf  (c)}  Outburst with a very steep rising 
branch  and  a sharp  peak. The  standard  deviations  of 
brightness are marked. The points denote the  1-day means 
of brightness connected by the line to guide the eye. The 
distances between the ticks on each axis are the same for 
all plots. See Sect.\,\ref{ana} for details.   } 
\label{exa}
\end{figure}

     To assess how mutually  different  the profiles  of the  light curves of 
the individual  outbursts  are, their  examples  from  Fig.\,\ref{ru-jd}a are 
displayed in  Fig.\,\ref{exa}. These events  differ  primarily  regarding the 
width, the rising branch's slope, and the slowly  decaying plateau's presence 
on the top. In a plateau on the outburst top, the  peak  was  associated with 
the plateau starts. A typical error of the determination of the maximum light 
time is 1--2\,days.

     The crossing  12.2\,mag  in  the  rising, and  the  outburst's  decaying 
branches were used to measure its length $L$. The  brightness  varied rapidly 
in these outburst  phases, which enabled  us to  determine  these  times with 
high accuracy (Fig.\,\ref{ru-jd}b). 

     The steepness of the rising branch of the  outburst was measured between 
the brightnesses of 12.0  and  11.0\,mag. The variety of the slopes was large 
here. The quantity  $\tau_{\rm  R}$  is expressed in days  for an increase in 
brightness of 1\,mag (Fig.\,\ref{ru-jd}c).    

     To investigate the  asymmetry of the light curve of an outburst, we used 
the parameter $P$ defined in Eq.\,\ref{eq1}):

\begin{equation}
\label{eq1}
P = \frac{T_{\rm peak} - T_{\rm start}}{T_{\rm end} - T_{\rm start}}     
\end{equation}

\noindent
where $T_{\rm  peak}$  denotes  the time of the peak of the outburst, $T_{\rm 
start}$  refers  to the  time  of the  start  of  this  event  (crossing  its 
brightness  12.2\,mag), and  $T_{\rm  end}$  refers  to the  time  of its end 
(crossing its brightness 12.2\,mag). These  quantities  were  determined from 
the light curve of each well-covered event. Their uncertainties were used for 
the calculation of the standard deviation of $P$. 

     Equation\,\ref{eq2})  defines  the  outburst  fluence  $S$ (the observed 
optical energy output of the individual outbursts) as:

\begin{equation}
\label{eq2}
  S = \int_{T_{\rm start}}^{T_{\rm end}} 10^{0.4 (12.2-V)} {\rm d}t
\end{equation}

\noindent
where $t$ is the time measured in days. Since we  are interested in comparing 
the  relative  outputs  of outbursts  in a  given DN,  $S$  was  expressed in 
dimensionless units (Fig.\,\ref{ru-jd}e).  


\begin{figure}
\centering
 \includegraphics[width=0.49\textwidth]{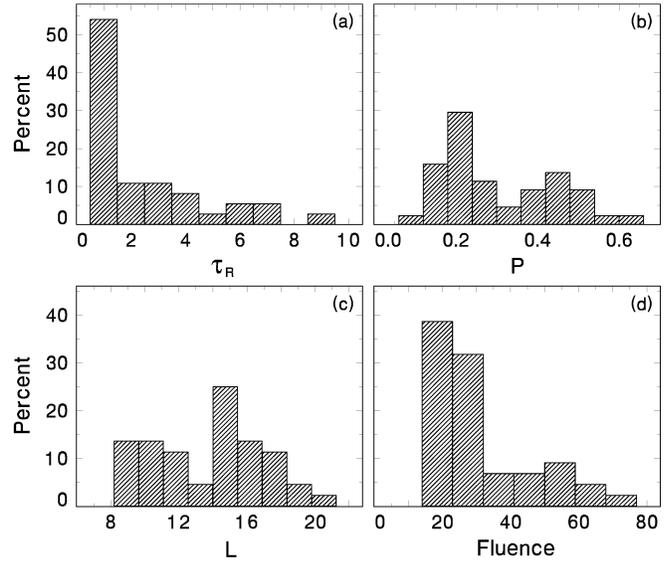} 
\caption{{\bf  (a)} Histogram of $\tau_{\rm  R}$ of the outburst 
in  RU\,Peg  from  Fig.\,\ref{ru-jd}c.  {\bf  (b)}  Histogram of 
parameter $P$ of the outburst from Fig.\,\ref{ru-jd}d. {\bf (c)} 
Histogram of length $L$ of the outburst from Fig.\,\ref{ru-jd}b. 
{\bf  (d)}   Histogram  of   fluences  of  the   outbursts  from 
Fig.\,\ref{ru-jd}e. See Sect.\,\ref{ana} for details.  } 
\label{hist}
\end{figure}

     In summary, out of the  52\,detected  outbursts, the reliable  values of 
$L$, $P$, and fluence  could be determined  for  44\,outbursts. The values of 
$\tau_{\rm  R}$  could be determined  for  37\,outbursts.  Figure\,\ref{hist} 
shows the  histograms  of  some  parameters  of the outbursts of RU\,Peg. The 
histogram of  $\tau_{\rm  R}$  is very  asymmetric, with  a tail toward  long 
$\tau_{\rm  R}$  (Fig.\,\ref{hist}a).   The  value  of  parameter  $P$ of the 
outburst spans a broad range, with two  mildly  defined peaks of  the  vastly 
different  heights  (Fig.\,\ref{hist}b).  Figure\,\ref{hist}c  shows  a  flat 
histogram of $L$ with very mildly defined peaks. A very  asymmetric histogram 
in Fig.\,\ref{hist}d shows an extensive range of fluences of the outbursts. 

\begin{figure}
\centering
 \includegraphics[width=0.49\textwidth]{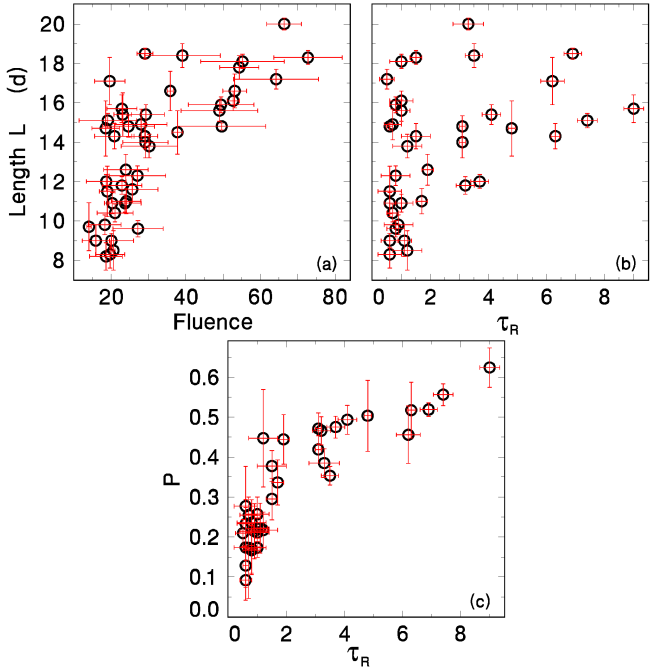} 
\caption{{\bf (a)} Relation of fluences and $L$ of outbursts 
in RU\,Peg. Notice the large scatter of $L$ for the fluences 
smaller than  30, followed  by  a long  tail  toward  bigger 
fluences.  {\bf  (b)} Relation of  $\tau_{\rm  R}$  and  $L$ 
of the outburst. {\bf  (c)}  Relation of $\tau_{\rm  R}$ and 
$P$ of the outburst.  See Sect.\,\ref{ana} for details.  } 
\label{flu-m}
\end{figure}

     A relation between the fluences and $L$ of the outbursts is displayed in 
Fig.\,\ref{flu-m}a. Notice  the  very  large  scatter of $L$ for the fluences 
smaller than 30 (where most outbursts accumulate), followed by a long tail of 
the longer outbursts  towards  higher fluences. Since points are dispersed in 
both the  direction of equal  $L$  and  equal fluences, variations of $L$ and 
outburst type contributed to the scatter. 

     Figure\,\ref{flu-m}b shows  that most  outbursts occupy a small range of 
$\tau_{\rm  R} < 2$\,d\,mag$^{-1}$, no matter how long these events are. Some 
outbursts with $L$ longer  than 12\,d  (and  especially  $>14$\,d)  can  have 
considerably bigger $\tau_{\rm  R}$. In Fig.\,\ref{flu-m}c, $\tau_{\rm R}$ is 
displayed vs $P$  taking  into account  the  whole  outburst. The relation of 
$\tau_{\rm  R}$ and $P$ shows a group of outbursts with small $\tau_{\rm R}$, 
sharply turning  into a gradual increase of $\tau_{\rm  R}$ of outbursts with 
$P>0.3$.

\subsection{Profiles of the outbursts    \label{types} }

\begin{figure}
\centering
 \includegraphics[width=0.49\textwidth]{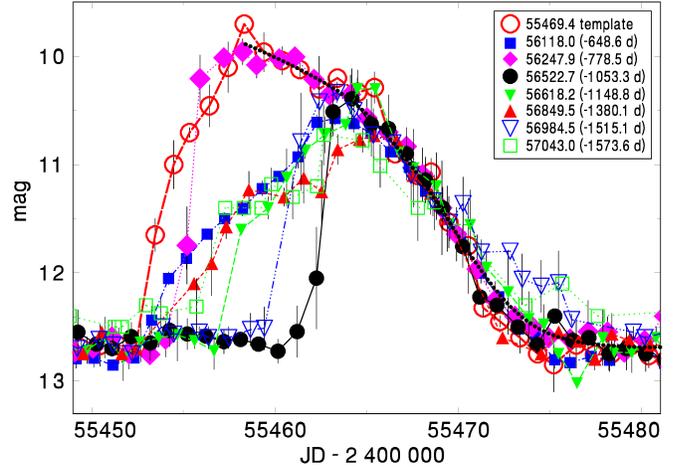} 
\caption{Light curves of the  well-observed  outbursts in RU\,Peg. 
The individual events  were  connected  by  the  line to guide the 
eye. The  error  bars  of  the  night  means  of  brightness  were 
marked. The  individual  events were shifted  along  the time axis 
to match the  decaying  branch  of  the  template.  The  times  of 
crossing 11.5\,mag on the decay, $t_{\rm d}$, (in JD--2\,400\,000) 
and the  shifts  (in  days)  are  listed. The  thick  dotted  line 
represents a fit to an ensemble  of  the decaying  branches of the 
outbursts. See Sect.\,\ref{ana} for details.  } 
\label{ru-dec}
\end{figure}

     The well-covered  outbursts  were  plotted and superposed. We found that 
the best match was obtained when these outbursts were co-aligned according to 
their decaying branches. One outburst with a well-covered decaying branch was 
chosen as the  template. The  remaining outbursts were shifted along the time 
axis to match this template's decaying branch (Fig.\,\ref{ru-dec}). The decay 
slopes remained mutually similar even if the  peak  position  with respect to 
the outburst's  centre  varied  appreciably. Even a primarily variable $L$ of 
the outburst did not affect the slope of the decay. 

     The decaying branches  of the  individual  outbursts  were merged into a 
common file  and  smoothed by  the code  HEC13, written  by Prof.~P.~Harmanec 
\footnote{\url{http://astro.troja.mff.cuni.cz/ftp/hec/HEC13/}}      (Harmanec 
1992). This code is based on  the  method of Vondr\'{a}k (1969), who improved 
the original method of  Whittaker (Whittaker \& Robinson 1946). It  can fit a 
smooth curve to the non-equidistant data no matter  what  their profile is. A 
full description can  be found in  Vondr\'{a}k (1969). This method enables us 
to find a compromise between a curve running  through all the observed points 
and an ideal smooth curve. 

     An ensemble of the best-covered  outbursts was used for a fitting of the 
decaying branches.  The input parameters of the fit $\epsilon = 10^{-1}$, the 
length of the bin $\Delta$$T = 0.05$\,d satisfied  the decaying branch of the 
ensemble. In our  case, the  input  parameters  were chosen  so that  the fit 
reproduced the decay's main profile (Fig.\,\ref{ru-dec}).

\subsection{The outburst recurrence time and its variations   \label{o-c} }

     Because the outbursts  of RU\,Peg  are  the  discrete events with easily 
resolvable peaks, the method  of the  O--C residuals of some reference period 
is suitable for an analysis of their $T_{\rm C}$ (`O' stands for observed and 
`C' for calculated times). Vogt  (1980)  applied this method to several DNe. 

     The resulting O--C  curve  also  enabled us to  assess  each  outburst's 
position with respect to the O--C profile  of  the remaining  outbursts. This 
method can  work  even if  some  outbursts  are  missing due to gaps in data, 
provided that the profile of the O--C curve is not  too complicated. 

    Minimizing the slope of the O--C values generated for various $T_{\rm C}$ 
yielded $T_{\rm  Cref}$, the  reference value of  $T_{\rm  C}$. The resulting 
ephemeris is given  in Eq.\,(\ref{eq3}). Time  of the basic peak of outburst, 
$T_{\rm b}$, is equal to JD\,2\,456\,237, ${T_{\rm Cref}}$ is equal to 90\,d, 
$E$ is epoch.

\begin{equation} 
t_{\rm max} = T_{\rm b}  + T_{\rm Cref} E  
\label{eq3} 
\end{equation}

\begin{figure} 
\centering
 \includegraphics[width=0.49\textwidth]{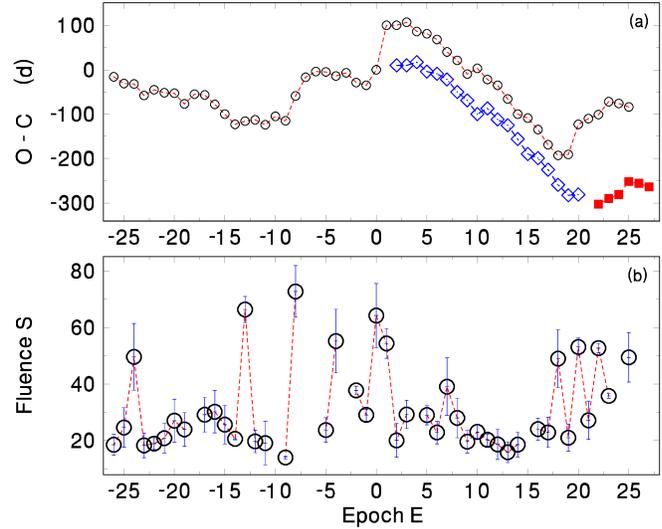} 
\caption{{\bf  (a)}  The  O--C  diagram  for  the  outbursts  in 
RU\,Peg. The open circles represent  the  O--C values calculated 
according to Eq.\,\ref{eq3} if we  assume that  we  observed all 
outbursts. The errors of the O--C values are 1--4\,d (comparable 
to the  sizes  of  the  symbols). The line  connects the  points 
considered  to  be the neighbouring outbursts. The open diamonds 
denote  the  O--C  curve of  nineteen\,outbursts  with  a missed 
outburst  between epochs  $E = 0$  and  $+1$. The filled squares 
denote  a segment  of the  curve  (six\,outbursts) with a missed 
outburst between $E = +19$ and $+20$. {\bf (b)} The evolution of 
fluences  of  the  outbursts  in  RU\,Peg  with  epoch  $E$ from 
Eq.\,\ref{eq3}.  A  line  connects  the  neighbouring  outbursts 
for  which  their  fluence  could  be  determined  reliably. See 
Sect.\,\ref{o-c} for details.     } 
\label{ru-oc}
\end{figure}

     Figure\,\ref{ru-oc}a shows the complicated  variations of the O--C curve 
around the mean value.  Their  amplitude  is considerably larger than that of 
the rapid outburst-to-outburst fluctuations.

\begin{figure}
\centering
 \includegraphics[width=0.47\textwidth]{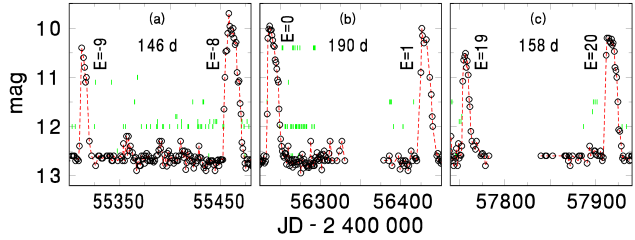} 
\caption{Light curves  of surroundings of the potentially 
missed outbursts in  RU\,Peg. Epochs of the two outbursts 
(Eq.\,\ref{eq3}) of each plot  and  their  separation  in 
days  are listed. The  points denote  the 1-day  means of 
brightness connected by the  line  to guide  the eye. See 
Sect.\,\ref{ana} for details.   } 
\label{m-out}
\end{figure}

     Also, modifications  of the  O--C  curve  for  several  possible  missed 
outbursts are shown in  Fig.\,\ref{ru-oc}a. A long gap between the  outbursts 
in the densely observed  light  curve in Fig.\,\ref{m-out}a shows that either 
no other outburst occurred between $E= -9$ and $E= -8$, or its peak magnitude 
was at least 2\,mag  lower  than  the surrounding  events. Inclusion  of  the 
missed  outbursts in seasonal gaps would alter the amplitude of  the jumps in 
the O--C curve (especially, in Fig.\,\ref{m-out}b).

     A comparison of the evolution of fluences with the O--C curve of $T_{\rm 
C}$ is shown in Fig.\,\ref{ru-oc}. The time segments of the relatively stable 
length of  $T_{\rm  C}$ (segments of $E$ between $-26$ and  $-15$ and between 
$+2$ and  $+17$)  were  accompanied  by  the  mostly roughly stable and small 
values of fluence. On the  contrary, the  series  of  the enormous changes of 
$T_{\rm  C}$  in other  time  segments  were  accompanied  by the significant 
variations of outburst fluences.

\section{Discussion}    \label{dis}    

     We present an analysis of the vigorous long-term optical activity of the 
DN RU\,Peg.

\subsection{The outburst types}    \label{typ}    

     We ascribe  outbursts  of RU\,Peg  shorter  than about 12\,d to be close 
to A-type (outside-in)  outbursts  because  of their  short  $\tau_{\rm  R} < 
2$\,d\,mag$^{-1}$ (Fig.\,\ref{flu-m}b).  In the  interpretation, while a bump 
in the  histogram  in  Fig.\,\ref{hist}c  with  shorter  $L$  outbursts  only 
contains the  probable  A-type  events  according  to their  $\tau_{\rm  R} < 
2$\,d\,mag$^{-1}$, the bump with $L$ longer  than  $12-14$\,d  consists  of a 
mixture of outbursts      of various  types because of their largely variable 
$\tau_{\rm  R}$  (up to  about  9\,d\,mag$^{-1}$). They  may start at various 
distances  from the  disc  centre,  $r_{\rm  s}$, and  span between the A and 
B-types (inside-out).  These outburst types were defined in the model of Smak 
(1984). 

     Bimodal histograms of $L$ of DNe  are  usual, and the prominence  of the 
peaks  differs  from  system  to  system  (Ak  et  al. 2002).  Although  this 
separation is  somewhat blurred in RU\,Peg, it also reflects the contribution 
of the different outburst types in RU\,Peg (Fig.\,\ref{hist}c).  

     Furthermore, outbursts of RU\,Peg occur in the restricted $\tau_{\rm R}$ 
vs $P$ region  (Fig.\,\ref{flu-m}c).  While  $\tau_{\rm  R}$  represents  the 
steepness of the rising branch, $P$ refers  to the asymmetry of the  outburst 
(given  primarily  by  the  whole  rising  branch  and  plateau  because  the 
decaying branches are mutually similar (Fig.\,\ref{ru-dec})). For example, no 
outbursts which may be close to the  $B$-type  contain  a significant plateau 
because  we  detect  no outbursts  in the region of  $\tau_{\rm  R} > 3$  and 
$P < 0.3$. In this context, the first bump in the histogram in Fig.\,\ref{hist}b 
represents outbursts similar to the A-type, the second bump in Fig.\,\ref{hist}b 
represents the tail which starts at $P\approx 0.35$ in Fig.\,\ref{flu-m}c and 
may contain  a mixture  of various  outburst types. It confirms the result in 
Fig.\,\ref{hist}c. 

     The rising branches in Fig.\,\ref{ru-dec} show a transition from the big 
initial steepness to the subsequent less  steep  rise in the different phases 
since the start for the individual outbursts (e.g., outbursts with $t_{\rm d} 
= 2\,456\,118$ and $t_{\rm  d} = 2\,456\,618$, both ascribed to be similar to 
the B-type). We interpret it as the outburst  starts  at the different values 
of $r_{\rm  s}$ in RU\,Peg. A discrepancy between outburst types and fluences 
can be explained by the dispersion of data for the fluences  smaller  than 30 
in both the direction of equal  $L$ and equal fluences in Fig.\,\ref{flu-m}a. 
It shows that  both  variations  of  $L$ and outburst type contributed to the 
scatter. A  detailed  analysis  of  the  rise  of  outbursts  (e.g.,  orbital 
modulation) and a comparison with models will be helpful. 

     As regards the differences  between  the outburst  types in RU\,Peg, the 
probable A-type outburst with $t_{\rm  d} = 2\,456\,522.7$,  $\tau_{\rm  R} = 
0.6$\,d\,mag$^{-1}$ was considerably  shorter  than the probable B-type event 
with  $t_{\rm  d} = 2\,456\,118.0$,  $\tau_{\rm   R}  =  6.9$\,d\,mag$^{-1}$) 
(Fig.\,\ref{ru-dec}). In the Smak (1984) model framework,  we  interpret this 
short likely  A-type  outburst  as starting  by reaching  the critical column 
density in a narrow ring at the outer disc rim. It is also possible to assess 
the values of  $r_{\rm  s}$  of the  future  events  by  comparing  models of 
relative delays  between  the optical  and  ultraviolet outburst light curves 
because outbursts  close to the  A-type have the  UV delay  considerably more 
prominent  than  the  B-types  (Smak  1998).  Also, the  changes  in  Doppler 
tomograms can shed more light; such a tomogram exists  only for  one outburst 
(Dunford et  al.  2012). Although  the  investigation of the evolution of the 
orbital  modulation  during  the  start  of  various   outbursts  of  RU\,Peg 
might  pinpoint  the  location  of the trigger of the instability, $i \approx 
33^{\circ}$ (Friend et al. 1990) indicates its low  amplitude and the absence 
of eclipses.

\subsection{Disc overflow}    \label{ov}    

     Although the B-type is typical for DNe with $\dot m_{\rm tr}$ lower than 
the A-type  (Smak 1984),  we interpret  the large variety of the light curves 
as the starts of outbursts of RU\,Peg at various values of $r_{\rm  s}$, even 
in roughly the same  $T_{\rm  C}$ (Figs.\,\ref{ru-jd} and \ref{ru-oc}a). Disc 
overflow may occur for some outbursts to modify  $r_{\rm  s}$  of the ring of 
the disc  matter  reaching  the  critical  column  density. It  gives rise to 
the accumulation of  matter in the different $r_{\rm  s}$  even  at a roughly 
constant value of $\dot  m_{\rm tr}$ in  the  model of Schreiber  \&  Hessman 
(1998). 

     This disc overflow, varying  with  time,  could also  contribute  to the 
change of  $\tau_{\rm  R}$  and  $P$. Doppler  tomography of  RU\,Peg  in one 
outburst (Dunford  et al. 2012)  shows  a highly  asymmetric extension of the 
width of the mass stream. Such tomograms  obtained in  various  outbursts and 
quiescences  could help assess  how and at which  places the collision of the 
inflowing matter  and  the  disc occurs at  various  times. Also, theoretical 
studies  of the effect of this merging on the outburst types and $T_{\rm  C}$ 
are desirable.

\subsection{Disc irradiation}    \label{rad}    

     We found that the properties of  the decaying  branches of the outbursts 
(hence the properties of  the  cooling  front  determining  the decay rate of 
these outbursts  (e.g., Smak  1984;  Hameury  et  al. 1998))  of RU\,Peg were 
stable  and  reproducible for the individual events, no  matter how steep the 
rising branches (whether close to the A or B-type) and how long the outbursts 
were (Fig.\,\ref{ru-dec}). The  profiles of the  outbursts' decays in RU\,Peg 
place the constraints  on the  strength  of irradiation  of the disc from the 
very hot WD (50\,000--70\,000\,K (Sion \& Urban 2002; Godon et al. 2008)) and 
the donor, as  modelled  by Hameury et  al. (1999, 2000). The  Hameury et al. 
(2000) model  predicts  several  noticeable  rebrightenings  on  the decaying 
branch of the outburst if the hot WD  strongly irradiates  the disc. However, 
these branches are smooth in  RU\,Peg, with  well-reproducing  profiles. This 
speaks in favour  of  no significant  irradiation  of the disc by the  WD. It 
could be reconciled by evaporation of the inner disc region in an outburst of 
RU\,Peg, modelled  by Hameury  et al. (1999). However, this evaporation would 
have to be strong enough no matter how big the fluence of the outburst was.  

     In this regard, the mutual similarity of the smooth decaying branches of 
the individual events in a given DN, irrespective of  the rising branches and 
the length of  $P_{\rm  orb}$, is  more general  (e.g., SS\,Cyg (Cannizzo  \& 
Mattei  1998),  DX\,And  (\v{S}imon 2000b), X\,Ser  (\v{S}imon 2018a)).  This 
ensemble also includes DO\,Dra (\v{S}imon 2000a) and GK\,Per (\v{S}imon 2002), 
which are  intermediate  polars  (IPs) (Haswell et al. 1997; Patterson et al. 
1992; Watson et al. 1985; Crampton  et al. 1986). IPs accrete  onto  the WD's 
polar caps in an outburst, not onto the  boundary  layer  (Warner  \&  McGraw 
1981; Warner 1995). All of  this speaks in favour of only weak irradiation of 
their discs, whether they accrete via  boundary  layer or polar caps. Because 
the propagation of  a cooling  front is not affected by irradiation  at large 
values of $r_{\rm s}$, we do not expect any difference between magnetised and 
non-magnetised  WD. Also, a low  $T_{\rm  eff}$ of the WD, evaporation of the 
inner  disc region or  a complicated  structure  of the inner disc region can 
contribute to preventing the disc from intense irradiation by the WD. 

   A flattening of the decay rate of most outbursts of RU\,Peg in their final 
phases, about 0.4\,mag  above  quiescence, can be explained by an increase of 
the mass  outflow  from  the donor's  nozzle  due  to its  irradiation in the 
advanced phase of the outburst (see the model of Hameury et al. 2000). Indeed, 
this irradiation in  outburst was  detected on the Roche tomograms of Dunford 
et al. (2012). Also, the light  contribution  of the secondary component, the 
WD and the hot spot  are  likely  to contribute  to this outburst phase. This 
behaviour is  different from  that in SS\,Cyg (Cannizzo  \&  Mattei 1998), in 
which the decay rate  of some  outbursts  displays a glitch about 0.8--1\,mag 
above quiescence, and  then  continues  with the  same  decay rate as before. 
Investigation of the evolution of the orbital modulation and the roles of the 
light contributions  of various system components of such DNe in these phases 
will be helpful.

\subsection{The variations of  $T_{\rm  C}$ }    \label{tc} 

     The variations of  $T_{\rm  C}$  between  the  neighbouring  (or nearby) 
outbursts were  much  smaller  than the  length  of  $T_{\rm  C}$ in RU\,Peg. 
Several episodes of the abrupt changes (jumps) were detected (Fig.\,\ref{ru-oc}a). 
All of this speaks in  favour  of mutually  dependent outbursts, although the 
fluence was  not  related  to the current length of  $T_{\rm  C}$. In the TVI 
model (Smak 1984; Hameury et al. 1998), only a small part of the disc  matter 
is accreted  from the disc in outbursts. Whether the outburst of RU\,Peg  was 
closer  to the  A-type  or  the  B-type and  how  big  $L$  and fluence  were 
(Figs.\,\ref{ru-jd} and  \ref{ru-oc}), it did not deviate from the O--C curve 
of other outbursts. The processes in the cool disc  thus played a significant 
role. This  curve  of RU\,Peg  is similar  to those  in other  DNe of various 
subtypes (e.g., Vogt 1980; \v{S}imon 2000b; \v{S}imon 2002). 

     A combination of the fluences with the evolution of  $T_{\rm  C}$ showed 
several time segments with  the  different  behaviours of  RU\,Peg.  The time 
segments with little variable  fluences, accompanied  by  a relatively stable 
length  of $T_{\rm  C}$, were sometimes replaced  by the time  segments  with 
a more remarkably variable  $T_{\rm  C}$  and scatter of fluences. The change 
reaching the peak-to-peak amplitude of the  scatter of fluences was sometimes 
achieved rapidly, even during the neighbouring outbursts (Fig.\,\ref{ru-oc}). 
A bigger fluence suggests  more  matter  accreted from the disc onto  the  WD 
during  this outburst  (e.g., Smak 1984; Hameury et al. 1998) irrespective of 
whether it was closer to the A or B-type. 

     The extremely large  fluences  of some  outbursts, occurring only in the 
events with excessively long $L$ ($>14$\,d) (Fig.\,\ref{flu-m}a) between $E = 
-14$ and $E = +2$ bear the similarity  to  a  very  long  outburst  of U\,Gem 
(Cannizzo et al. 2002), attributed to an  increase in the  mass transfer rate 
from the donor to the disc during the DN outburst. In the interpretation, the 
activity of the  donor in  RU\,Peg  (e.g., an  increase  of the  mass outflow 
through the  L$_{1}$  point)  caused  considerable variations of  fluence. We 
attribute it to destabilising the disc and  $T_{\rm  C}$ without any definite 
correlation in this time segment.

     Also, a prominent jump  near  $E = +18$  is an example that the relation 
between $\dot m_{\rm tr}$ and $T_{\rm C}$ of RU\,Peg can be exceeded by other 
processes. A model with a variable outer disc boundary (Hameury  et al. 1998) 
shows that an increase of  $\dot  m_{\rm  tr}$  by a factor of ten leads to a 
shortening of $T_{\rm C}$ by about 50\% and broader, more luminous outbursts. 
Although the jump near $E = +18$ gives an increase of  $T_{\rm  C}$, a series 
of the large variations of the outburst fluences (with a significant fraction 
of outbursts  with  large  fluences, hence  increases  of the  amount  of the 
accreted matter) started in the vicinity of this jump (Fig.\,\ref{ru-oc}). We 
ascribe the decrease of  $\tau_{\rm  R}$  and $P$ since  the vicinity of $E = 
+18$  to  the  dominance  of  the  A-type events, accompanying this change of 
$T_{\rm  C}$  (Figs.\,\ref{ru-jd}c and \ref{ru-jd}d). We ascribe this episode 
to a change of the disc structure. 

     Rapid variations of $T_{\rm C}$ can also be caused by a variable removal 
of angular momentum  from  the  matter  in  the  disc, for  example,  if  the 
magnetic field from the starspots  on the  secondary reaches over to the disc 
(Meyer-Hofmeister et al. 1996). In this model, the  interaction of this field 
with the disc leads to shorter DN outbursts and a decrease of $T_{\rm C}$ for 
a given mass transfer rate between the components. The time segments in which 
the effect of this  field on  the  disc  dominates (e.g., no bursts of matter 
from the donor occur)  might explain  the relatively stable and short $T_{\rm 
C}$, accompanied  by the  small  fluences  and  the  often  small  $L$ in $E$ 
between $-23$ and $-15$  and  especially  between  $+2$ and  $+17$ in RU\,Peg 
(Fig.\,\ref{ru-oc}a).   

     In conclusion, a combination  of the variations of $T_{\rm  C}$ with the 
complicated and unstable  properties of  the outbursts, including an unstable 
mass transfer rate between the components, suggest the  dominance  of several 
mechanisms on the state of the disc of RU\,Peg in various time segments.

\section*{Acknowledgments}
This study was supported by the EU project H2020 AHEAD2020, grant agreement 
871158. Also, support  by  the project  RVO:67985815 is  acknowledged. This 
research used the observations from the AAVSO International database  (USA) 
and  the  AFOEV  database  (France). I thank the  variable  star  observers 
worldwide. I also  thank Prof.~Petr~Harmanec for providing me with the code 
HEC13. The Fortran source version, compiled version, and brief instructions 
on how to use the program can be obtained at: 
http://astro.troja.mff.cuni.cz/ftp/hec/HEC13/

\section*{Data availability}
The dataset was derived from a source in the public domain.

\label{lastpage}

\end{document}